\documentclass[prx,twocolumn,english,
,groupedaddress,
floatfix,longbibliography]{revtex4-2}

\pdfoutput=1
\usepackage[utf8]{inputenc}
\usepackage[T1]{fontenc}
\usepackage{braket,xcolor}
\usepackage{graphicx,enumerate,verbatim,bbold}
\usepackage{amsmath,amssymb,amsthm,float,mathrsfs}
\usepackage{dsfont}
\usepackage[normalem]{ulem}
\usepackage{hyperref}
\hypersetup{colorlinks,citecolor=blue,linkcolor=blue,urlcolor=blue}
\usepackage{braket}
\usepackage{multirow}
\usepackage{bbm}
\usepackage{physics}
\usepackage{bm}
\usepackage{tikz}
\usetikzlibrary{patterns}
\usetikzlibrary{fadings}
\usetikzlibrary{quantikz2}
\usepackage[normalem]{ulem}
\usepackage[caption=false]{subfig}
\usepackage{dcolumn}
\allowdisplaybreaks

\newcommand{\bqa}{\begin{eqnarray}}
\newcommand{\eqa}{\end{eqnarray}}

\newcommand{\be}{\begin{equation}}
\newcommand{\ee}{\end{equation}}

\newcommand{\del}[1]{\hil{{\bf XXX}}}

\newcommand{\hil}[1]{{\color[rgb]{0.6,0,0}{#1}}}
\newcommand{\um}{\mathbbm{1}}
\usepackage{subfig}

\definecolor{nqdcolor}{rgb}{0.5586, 0.0586, 0.4219}

\usepackage{comment}

\newcommand{\Imperial}{Blackett Laboratory, Imperial College London, SW7 2AZ, United Kingdom}

\begin{document}
\title{
Accurate and bounded approximation of quantum correlation functions
}


\author{Alexander van Lomwel}\affiliation{\Imperial}
\author{Florian Mintert}\affiliation{\Imperial}


\begin{abstract}
The exact classical simulation of non-integrable quantum systems rapidly becomes intractable beyond small system sizes, yet its role is central to predicting many-body dynamics. 
Approximate methods alleviate this cost but typically lack rigorous guarantees on their deviation from the exact result. 
For the infinite-temperature correlation function, we derive a rigorous error bound for approximating the full dynamics within a computable finite-sized region. 
Applied to autocorrelations, this finite-region approach is especially accurate across one-dimensional systems and beyond, and can further provide rigorous error bounds for other approximate methods, demonstrated here for matrix-product-operator simulations.
\end{abstract} 
\maketitle

\section{Introduction}
The simulation of quantum dynamics by classical means provides a crucial route to predicting and interpreting quantum phenomena~\cite{georgescu2014quantum,altman2021quantum}. 
Of particular interest is the simulation of non-integrable quantum systems, which permit the study of nonequilibrium quantum dynamics and quantum-chaotic behavior~\cite{rigol2009breakdown,hahn2024eigenstate}.
Their dynamics involve phenomena including thermalization and prethermalization~\cite{rigol2009breakdown,hahn2024eigenstate,mallayya2019prethermalization}, diffusive and anomalous transport~\cite{mcroberts2024parametrically}, the spreading and scrambling of quantum information~\cite{xu2019locality,parker2019universal}, and
atypical dynamics through mechanisms such as many-body localization and quantum many-body scarring~\cite{abanin2019colloquium,kerschbaumer2025quantum}.

For generic interacting systems, however, exact simulation is restricted to small system sizes.
Important exceptions arise in specially structured systems, including models whose dynamics can be reduced to non-interacting fermions~\cite{wiersema2024classification,vona2025exact}, or confined to polynomially scaling Lie algebras~\cite{wiersema2024classification,orozco2024quantum,van2026fast,van2026quantum}.
For generic non-integrable systems, one therefore typically relies on approximate methods that enable simulations with given, limited computational resources.

In order to distinguish an actual physical phenomenon from an artefact of an approximation, the ability to assess the accuracy of an approximation is essential.
Substantial work has thus gone into the derivation of bounds to the deviation between approximations and exact dynamics.
Such bounds exist for some common approximations such as Trotter decompositions~\cite{burgarth2024strong,hahn2025lower} and variational quantum time evolution~\cite{zoufal2023error}, but there is also a range of frequently used approximations whose accuracy can not be bounded yet.

The implications of the lack of such bounds
are particularly evident in matrix-product-state (MPS) and matrix-product-operator (MPO) simulations, which are amongst the most powerful tools for approximating one-dimensional dynamics~\cite{haegeman2016unifying,hemery2019matrix,le2026mitigating}.
Their efficiency
depends on representing the evolving state or operator with a tractable bond dimension, but the rapid growth of state or operator entanglement can require a rapidly increasing bond dimension, such that finite-bond dimension simulations may deviate substantially from the exact dynamics~\cite{hemery2019matrix,zhou2017operator,grundner2024complex}.
While convergence with increasing bond dimension indicates numerical stability, 
it does not necessarily guarantee agreement with the exact dynamics~\cite{kloss2018time}.

A particularly important target for quantum simulation are dynamical correlation functions.
By quantifying correlations between observables evaluated at different times, these functions serve as vital diagnostics of quantum many-body dynamics.
For a given initial state, their variation in time can show how correlations spread through the system following a quantum quench~\cite{calabrese2006time}, relaxation in isolated many-body systems~\cite{alhambra2020time}, and finite-temperature dynamical response and transport~\cite{kokalj2009finite,bertini2021finite}.
A particularly useful limit is provided by infinite-temperature dynamical correlations, for which all states in the Hilbert space contribute with equal statistical weight.
This thus provides a route to study how an initially local observable spreads throughout the system across the full energy spectrum, and has been used to study diffusive and anomalous transport~\cite{richter2019magnetization,shi2024probing}, hydrodynamic scaling~\cite{ljubotina2019kardar}, and nonequilibrium density profiles and steady states from closed-system dynamics~\cite{heitmann2023spin,kraft2024lindblad}.

In this work, we build on locality-based descriptions of operator growth, whereby the influence of a local operator remains approximately confined to a finite region for finite evolution times~\cite{bravyi2006lieb,wang2021bounding}.
We show that correlation functions can be approximated in terms of local dynamics with substantially higher accuracy than the underlying observables,
and we provide an exactly evaluable bound for the local approximation of correlation functions that is consistent with this increased accuracy.
The framework is exemplified for the autocorrelation function~\cite{udupa2023weak,yeh2023decay,vernier2024strong,yin2021prethermal,teretenkov2025pseudomode} for one- and two-dimensional non-integrable systems, and is extended to provide bounds for other approximate methods, with the explicit example of MPO simulation.

\section{Window approximation \& bounds} 

\begin{figure}[t]
\centering    \includegraphics[width=\linewidth]{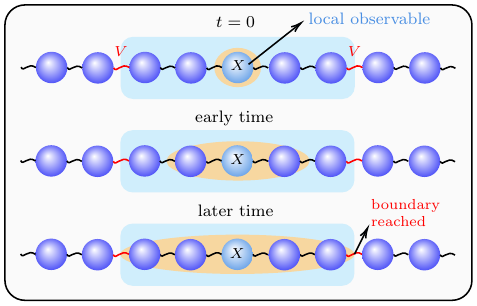}
\caption{
A schematic sketch of the window approximation, exemplified for a one-dimensional system.
    The local observable $X$ is initially confined to the central site of a finite-window of spins (shown by the blue region) within the larger system.
    As this observable evolves with the system's interactions, shown by the black connecting lines, the observable expands to operators with support growing from the initial site (shown by the growing orange region).
    After a finite time, the operator growth reaches the window boundary couplings $V$, shown in red, where the evolution of $X$ within the window can deviate from the full-system evolution.}
    \label{fig:schematic_sketch}
\end{figure}

Given a time-dependent Hamiltonian describing an interacting quantum spin system, and a finite window of spins within that system, the full-system Hamiltonian $H_F(t)$ can be decomposed into three parts: $H_F(t)=H_W(t) + H_\text{out}(t) + V(t)$,  where $H_W(t)$ is the Hamiltonian with support only on that window, $H_\text{out}(t)$ contains all the terms acting entirely outside the window, and $V(t)$ 
includes the terms that couple the window to the rest of the system.
With an observable $A$ supported entirely inside the spin window, the observable evolution is given by
\begin{equation}
    A_j(t,s) = U^\dagger_j(t,s) \ A \ U_j(t,s)
    \label{eq:Aevolution}
\end{equation}
from an initial time $s$ to a final time $t$, in terms of the propagator $U_j(t,s)$ induced by the Hamiltonian $H_j(t)$, with $j=F$ for the full-system evolution and $j=W$ for evolution exclusively within the finite-window. 
The following adopts the notation $A_j(t,0)\equiv A_j(t)$.
While the observable within the window is evolved with the Hamiltonian $H_W(t)$ instead of the full Hamiltonian $H_F(t)$, for sufficiently short times, the operator $A_W(t)$, understood on the full system as $A_W(t)\otimes \um_\text{out}$, can closely approximate the exact evolution $A_F(t)$ at sufficiently short times.
This follows from the locality of operator growth, where the window approximation is exact to a given order in perturbation theory, {\it i.e.}
as long as the growth of $A$ remains contained within the window, the window and full evolutions are effectively identical. 
Deviations arise only once the operator reaches the window boundary and the omitted couplings to the exterior begin to influence its evolution.
A schematic sketch of this framework, illustrated for a one-dimensional system, is shown in Fig.~\ref{fig:schematic_sketch}.

The following is based on applying this framework to the infinite-temperature dynamical correlation function,
\begin{equation}
    C_j(t)= 2^{-n}\tr (A_j(t)B) \ ,
    \label{eq:auto_generic}
\end{equation}
where, as with $A$, $B$ is an observable supported entirely inside the window.
Although Eq.~\eqref{eq:auto_generic} explicitly evolves only $A$, the accuracy of the window correlation, $C_W(t)$, is governed by the operator growth of both $A$ and $B$.
As shown in Sec.~\ref{sec:approxCorr}, the deviation of the correlation function $\Delta C(t)=C_F(t) - C_W(t)$ grows at least with the order $t^{m_A + m_B}$, where $t^{m_A}$ and $t^{m_B}$ are the time orders for which the evolution of the observables $A$ and $B$ within the window deviate from the full-system dynamics.
The correlation function can therefore remain accurately approximated by the finite window well beyond the time over which either evolved observable is itself accurately approximated.
Consequently, any bound to the observable deviation is generally too conservative to be used as a bound to the correlation deviation. 
In Sec.~\ref{sec:bounds}, we thus derive a bound specifically for the correlation function (Eq.~\eqref{eq:auto_generic}).

\subsection{Equations of motion}
\label{sec:eqom}

To derive the order at which $\Delta C(t)$ grows,
it is helpful to first outline the relevant equations of motion.
The correlation function (Eq.~\eqref{eq:auto_generic}) can be equivalently written in terms of the normalized trace $2^{-n}\tr (AB_j(t))$, with
\begin{equation}
    B_j(t,s) = U_j(t,s) \ B \ U^\dagger_j(t,s) \ ,
    \label{eq:Bevolution}
\end{equation}
and $B_j(t,0)\equiv B_j(t)$.
While the operator $B_j(t,s)$ satisfies the standard von Neumann equation of motion 
\begin{equation}
    \frac{\partial B_j(t,s)}{\partial t} = -i[H_j(t), B_j(t,s)] \ ,
    \label{eq:Beqom}
\end{equation}
with the Hamiltonian $H_j$ that induces the propagator $U_j$, the operator $A_j(t,s)$ (Eq.~\eqref{eq:Aevolution}) satisfies the von Neumann equation 
\begin{equation}
    \frac{\partial A_j(t,s)}{\partial t}= i[U^\dagger_j(t,s) \ H_j(t) \ U_j(t,s), A_j(t,s)] \ ,
\end{equation}
with a more complicated Hamiltonian.
In the case of a time-independent Hamiltonian, the equality $U_j^\dagger H_jU_j=H_j$ holds, since the propagator $U_j$ commutes with its generator $H_j$; but in the case of a time-dependent Hamiltonian, this equality does typically not hold.

To circumvent the technical difficulties resultant from the more complicated Hamiltonian, it is helpful to consider the equation of motion in terms of the initial time $s$,
\begin{equation}
    \frac{\partial A_j(t,s)}{\partial s}= -i[H_j(s), A_j(t,s)] \ ,
    \label{eq:Aeqom}
\end{equation}
with the actual Hamiltonian $H_j$.

Using Eqs.~\eqref{eq:Beqom} and \eqref{eq:Aeqom}, equations of motion of the observable deviations $\Delta A(t,s)= A_F(t,s) - A_W(t,s)$ and $\Delta B(t,s) = B_F(t,s) - B_W(t,s)$ are thus given by
\begin{equation}
    \frac{\partial \Delta A(t,s)}{\partial s} = -i[H_F(s), \Delta A(t,s)] - i[V(s), A_W(t,s)] \ ,
    \label{eq:DAeqom}
\end{equation}
in terms of the initial time $s$, and
\begin{equation}
    \frac{\partial \Delta  B(t,s)}{\partial t} = -i[H_F(t),\Delta B(t,s)] - i[V(t), B_W(t,s)] \ ,
    \label{eq:DBeqom}
\end{equation}
in terms of the final time $t$, respectively.

The solutions to these equations of motion are
\begin{equation}
    \Delta A(t,s) = i\int^t_s dr \ U_F^\dagger(r,s) [V(r), A_W(t,r)] U_F(r,s) \ ,
    \label{eq:DAsolution}
\end{equation}
and
\begin{equation}
    \Delta B(t,s) = -i \int^t_s dr \ U_F(t,r) [V(r),B_W(r,s)] U^\dagger_F(t,r) \ ,
    \label{eq:DBsolution}
\end{equation}
respectively, where, in the following, $\Delta A(t,0)\equiv \Delta A(t)$ and $\Delta B(t,0)\equiv \Delta B(t)$.

\subsection{Accuracy of the window approximation}
\label{sec:approxCorr}

The equations of motion outlined in Sec.~\ref{sec:eqom} can be used to derive an exact expression for the correlation deviation $\Delta C(t)$, from which the order in time at which the deviation grows can be extracted.
With the correlation deviation expressed as $\Delta C(t) = 2^{-n}\tr (\Delta A(t)B)$, substituting Eq.~\eqref{eq:DAeqom} yields the expression,
\begin{align}
    \Delta C(t) &= i2^{-n} \int^t_0 dr \tr (U_F^\dagger(r,0) [V(r), A_W(t,r)] U_F(r,0)B) \notag \\
    &=i2^{-n} \int^t_0 dr \tr([V(r), A_W(t,r)] B_F(r))
    \label{eq:DCmidstep}
\end{align}
using the cyclicity of the trace.
Under the condition $\tr_\text{out}V=0$, which is automatically satisfied for standard traceless Pauli couplings across the boundary, the term $\tr([V,A_W]B_W)$ equals zero, which permits Eq.~\eqref{eq:DCmidstep} to be written as
\begin{align}
    \Delta C(t) &= i2^{-n} \int^t_0 dr \tr([V(r), A_W(t,r)] (B_F(r) - B_W(r))) \notag \\
    &= i2^{-n} \int^t_0 dr \tr([V(r), A_W(t,r)] \Delta B(r)) \ .
    \label{eq:DCexact}
\end{align}

Given local observables, the bound $\Delta C(t)$ grows with a given power in $t$ that depends on the powers $m_A$ and $m_B$ that determine the accuracy of the window approximation of the observables $A$ and $B$.
The factor $\Delta B(r)$ grows as $r^{m_B}$.
In order to identify the growth of the factor $[V(r), A_W(t,r)]$ in Eq.~\eqref{eq:DCexact}, it is helpful to notice that this term also appears in Eq.~\eqref{eq:DAeqom}.
Since $\Delta A(t,r)$ grows as $(t-r)^{m_A}$, the first derivative of this quantity necessarily grows as $(t-r)^{m_A-1}$.
Thus, the right-hand-side of Eq.~\eqref{eq:DAeqom} additionally grows as $(t-r)^{m_A-1}$.
The first term $[H_F(r), \Delta A(t,r)]$, however, grows as $(t-r)^{m_A}$.
Thus, since Eq.~\eqref{eq:DAeqom} holds for all values of $t$, the term $[V(r), A_W(t,r)]$ must grow as $(t-r)^{m_A-1}$.
With the two factors in Eq.~\eqref{eq:DCexact} growing as $(t-r)^{m_A-1}$ and $r^{m_B}$, and the additional integration, the deviation $\Delta C(t)$ thus grows at least as $t^{m_A+m_B}$.

\subsection{Error bounds}
\label{sec:bounds}

The exact deviations derived in Sec.~\ref{sec:approxCorr} (Eq.~\eqref{eq:DAsolution}, Eq.~\eqref{eq:DBsolution}, and Eq.~\eqref{eq:DCexact}) expectedly depend on full-system quantities that are generally unavailable.
The goal is to thus derive useful bounds to the deviation errors that can be evaluated with quantities only within the finite-window.

The formal solution (Eq.~\eqref{eq:DAsolution}) for the difference $\Delta A(t,s)$ of exact dynamics and window approximation yields the bound~\cite{wang2021bounding}
\begin{equation}
    \norm{\Delta A(t)} \le \int^t_0 dr \ \norm{[V(r), A_W(t,r)]} =: \mathcal{B}_O(t) \ ,
    \label{eq:operatorbound}
\end{equation}
in terms of the normalized Hilbert-Schmidt norm $\norm{X} = \sqrt{2^{-n}\tr (X^\dagger X)}$, which can be evaluated without the full-system evolution.

Using the expression Eq.~\eqref{eq:DCmidstep}, and the Cauchy-Schwarz inequality, $\abs{2^{-n}\tr (X^\dagger Y)}\le \norm{X}\norm{Y}$, one has the bound
\begin{equation}
    \abs{\Delta C(t)} \le \int^t_0 dr \norm{[V(r), A_W(t,r)]} \norm{B_F(r)} \ .
\end{equation}
Given that $\norm{B_F(r)} = \norm{B}$, and for any Pauli observable $B$ which necessarily has unit normalized norm, the bound $\mathcal{B}_O(t)$ (Eq.~\eqref{eq:operatorbound}) is additionally a valid bound for the deviation $\abs{\Delta C(t)}$.
As determined in Sec.~\ref{sec:approxCorr}, the factor $[V(r), A_W(t,r)]$ grows as $(t-r)^{m_A-1}$.
With the integration, this bound thus grows as $t^{m_A}$. 
Given that the deviation $\abs{\Delta C(t)}$ grows at least as $t^{m_A+m_B}$, the bound $\mathcal{B}_O(t)$
is in general overly conservative.

The goal is to thus derive a stronger bound specifically for this deviation. 
Using the exact expression for $\Delta C(t)$ (Eq.~\eqref{eq:DCexact}) and the Cauchy-Schwarz inequality,
one has
\begin{equation}
    \abs{\Delta C(t)} \le \int^t_0 dr \norm{[V(r), A_W(t,r)]} \norm{\Delta B(r)} \ .
    \label{eq:DCbound}
\end{equation}
With the exact expression for the deviation $\Delta B(t,s)$ (Eq.~\eqref{eq:DBsolution}), this can be bounded similarly to Eq.~\eqref{eq:operatorbound}, which permits the replacement in Eq.~\eqref{eq:DCbound},
\begin{equation}
    \norm{\Delta B(r)} \le \int^r_0 dp \norm{[V(p),B_W(p)]}
\end{equation}
to yield the bound $\abs{\Delta C(t)}\le \mathcal{B}_C(t)$ for the correlation deviation, with
\begin{equation}
 \mathcal{B}_C(t) = \int^t_0 dr \norm{[V(r), A_W(t,r)]}\int^r_0 dp \norm{[V(p),B_W(p)]}
\label{eq:correlationbound}
\end{equation}
which can be evaluated without any full-system quantities.
The factor $[V(p),B_W(p)]$ grows as $p^{m_B-1}$, and with the inner integration, this term then grows as $r^{m_B}$.
With the factor $[V(r), A_W(t,r)]$ growing as $(t-r)^{m_A-1}$, and together with the growth $r^{m_B}$, the final integration over $r$ yields the bound growth $t^{m_A+m_B}$. 
This scaling is therefore consistent with the exact behavior of $\Delta C(t)$, making $\mathcal{B}_C(t)$ a considerably tighter bound on the correlation deviation than $\mathcal{B}_O(t)$ (Eq.~\eqref{eq:operatorbound}).

Notably, the bounds $\mathcal{B}_O(t)$ (Eq.~\eqref{eq:operatorbound}) and $\mathcal{B}_C(t)$ (Eq.~\eqref{eq:correlationbound}) depend only on the finite window and its boundary interactions, but not on the total system size.
These bounds thus remain valid as the surrounding system is enlarged and thus extend naturally to the thermodynamic limit.

\section{Results}

\label{sec:1D}
\begin{figure}[t]
\centering
\includegraphics[width=\linewidth]{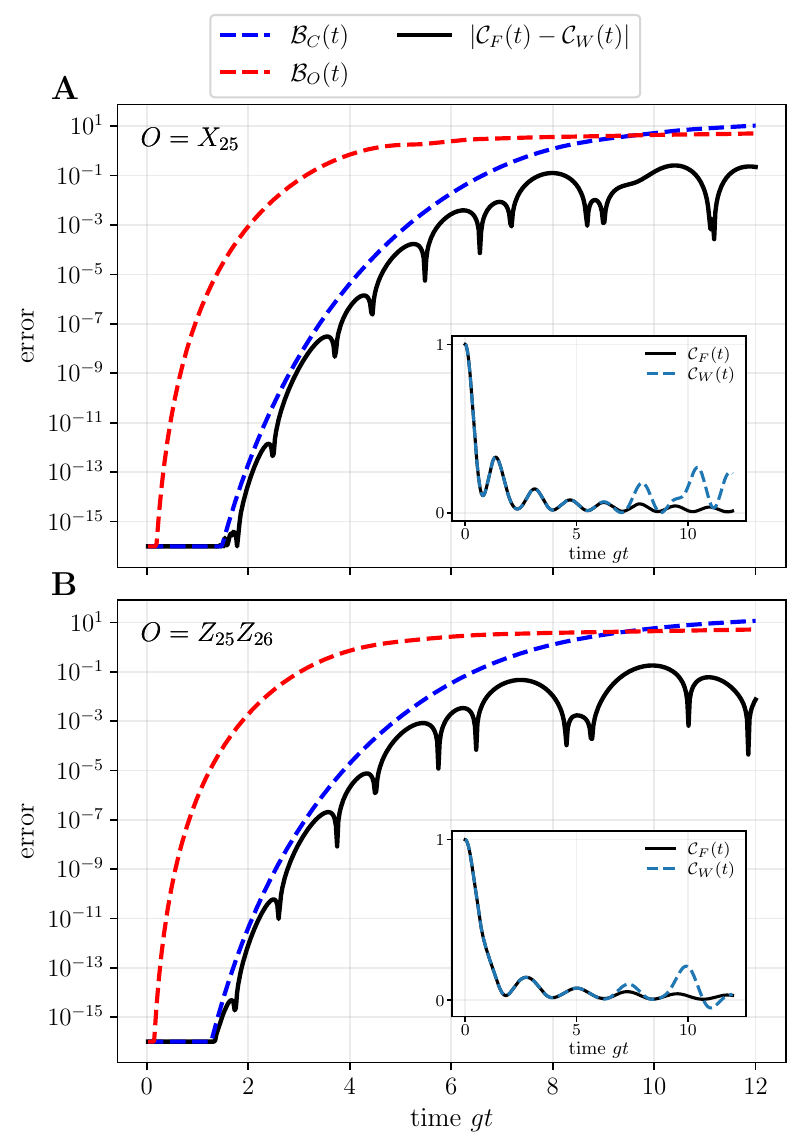}
\caption{
Example for the integrable transverse-field Ising model ({\it i.e.} $h_z=0$ in Eq.~\eqref{eq:MFIM}).
The autocorrelation function $\mathcal{C}_F(t)$ (Eq.~\eqref{eq:auto_function}) corresponds to a chain of $n=50$ spins, and is approximated by the finite-window autocorrelation function $\mathcal{C}_W(t)$ (Eq.~\eqref{eq:auto_function}) for a $13$-spin window centered at site $j_0=25$.
Plotted are bounds to the autocorrelation error in terms of the observable bound  $\mathcal{B}_O(t)$ (Eq.~\eqref{eq:operatorbound}) in red, and the correlation bound $\mathcal{B}_C(t)$ (Eq.~\eqref{eq:correlationbound}) in blue, as well as the true error computed via the Lie algebra of the integrable transverse-field Ising model, as functions of time $gt$, where $g$ is the Hamiltonian's interaction strength.
Panel {\bf A} corresponds to observable $O=X_{j_0}$ and panel {\bf B} corresponds to $O=Z_{j_0}Z_{j_0+1}$.
In both cases, $B_C(t)$ provides a substantially improved bound.
The bottom right insets show $\mathcal{C}_F(t)$ and $\mathcal{C}_W(t)$ as functions of time.
}
    \label{fig:exact_auto}
\end{figure}

\begin{figure}[t]
    \centering
    \includegraphics[width=\linewidth]{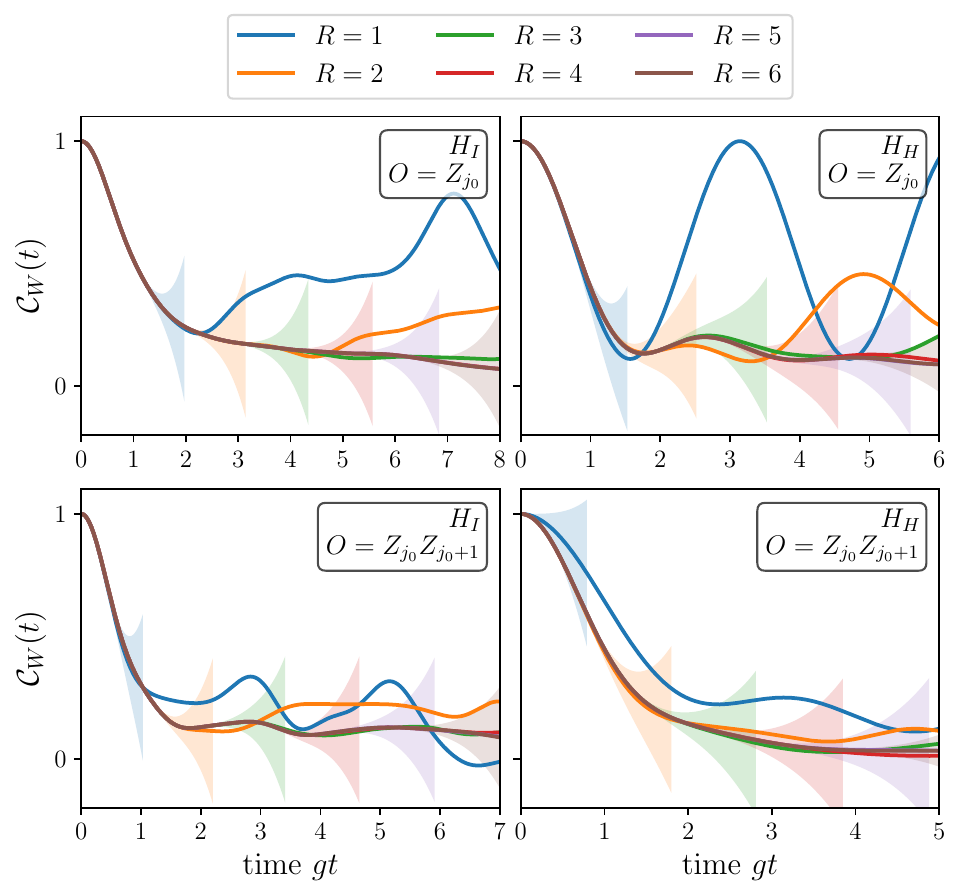}
    \caption{The window approximated autocorrelation function $\mathcal{C}_W(t)$ (Eq.~\eqref{eq:auto_function}) for window radii $R\in[1,6]$, as a function of time $gt$, in terms of the interaction strengths $g$ of the Hamiltonians $H_I$ and $H_H$.
    The data is for the non-integrable Hamiltonian $H_I$ (mixed-field Ising model, Eq.~\eqref{eq:MFIM}) and $H_H$ (Heisenberg model, Eq.~\eqref{eq:XXX}), as well as the single-site observable $O=Z_{j_0}$ and two-site observable $O=Z_{j_0}Z_{j_0+1}$, with $j_0$ as the central window site.
    The full-system function $\mathcal{C}_F(t)$ is guaranteed to be within the shaded regions  $\mathcal{C}_W(t)\pm \mathcal{B}_C(t)$, with the correlation bound $\mathcal{B}_C(t)$ (Eq.~\eqref{eq:correlationbound}).
    For early times, the function $\mathcal{C}_W(t)$ coincides for all values of $R$, until functions with smaller $R$ begin to deviate.
    As $R$ is increased, the error region, shown only for $\mathcal{B}_C(t)\le 0.3$, is delayed to later times.
    For $R=6$, this thus provides a substantial time interval for which $\mathcal{C}_W(t)$ closely approximates $\mathcal{C}_F(t)$.}
    \label{fig:error_regions}
\end{figure}

\begin{figure}[t]
    \centering
    \includegraphics[width=\linewidth]{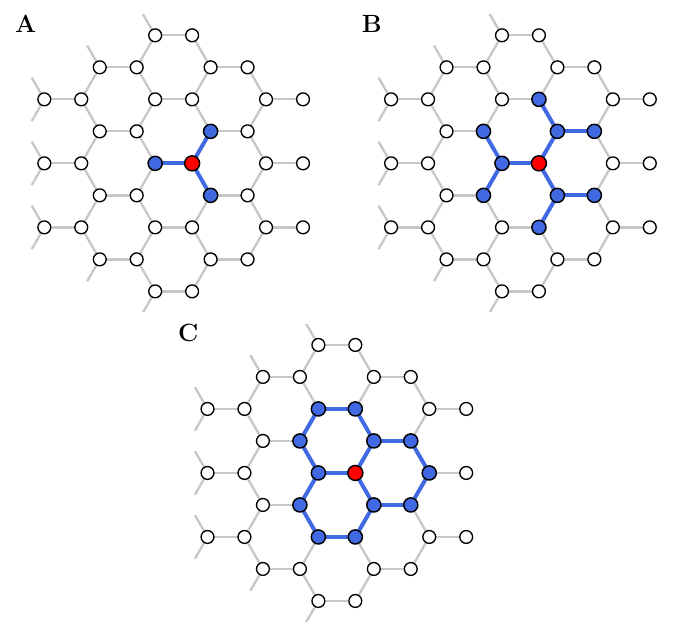}
    \caption{Window configurations for the honeycomb model, with the red spin denoting the initial observable location $j_0$. 
    Configuration {\bf A} extends one spin outward from the center, while configuration {\bf B} extends two spins outward. 
    Configuration {\bf C} is obtained by adding three spins to configuration {\bf B}, completing three full hexagons.
    }
    \label{fig:honeycomb}
\end{figure}

\begin{figure*}[t]
    \centering
    \includegraphics[width=\linewidth]{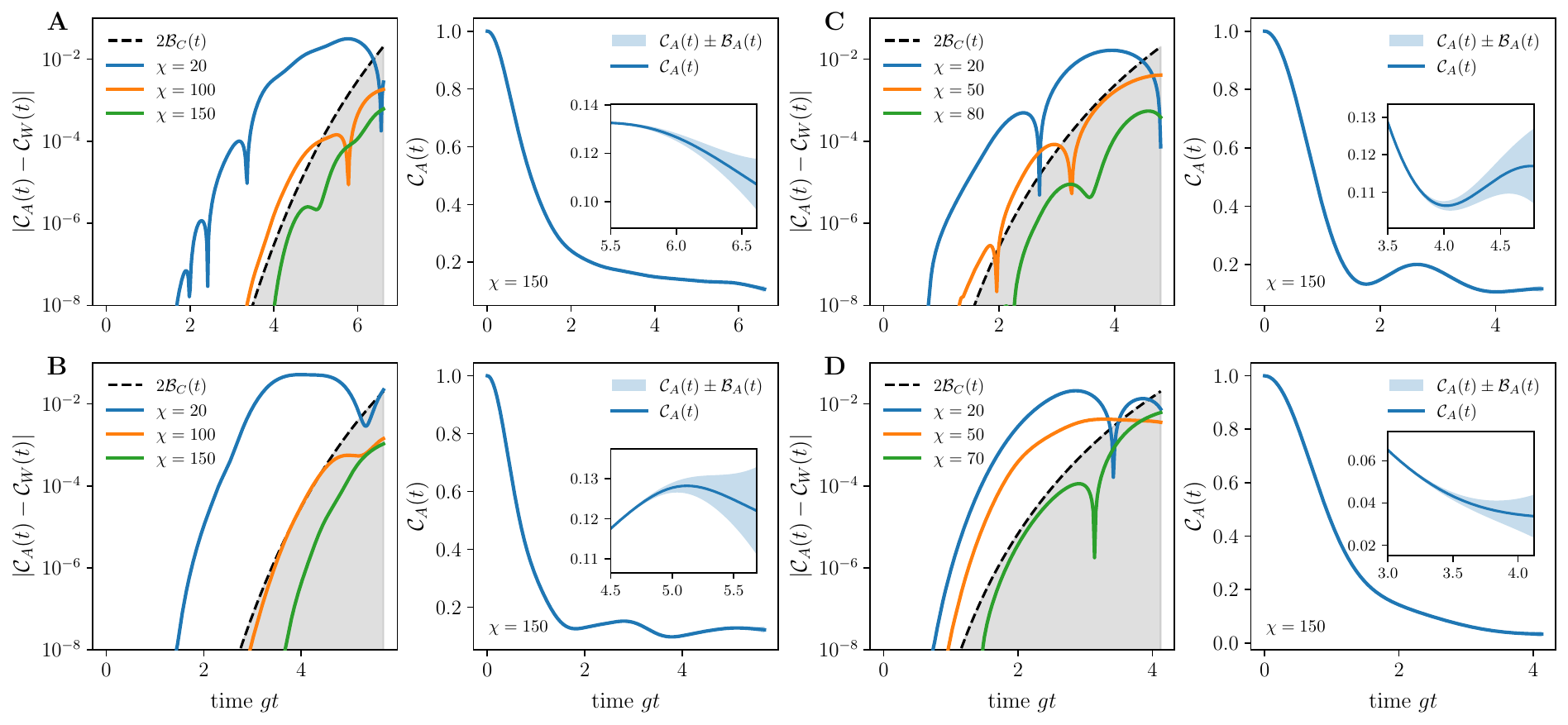}
    \caption{MPO simulation of the autocorrelation function $\mathcal{C}_A(t)$ to approximate the full-system function $\mathcal{C}_F(t)$ (Eq.~\eqref{eq:auto_function}) for a chain of $n=51$ spins with bond dimension $\chi$, using TeNPy fourth order TEBD with time-step $dt=0.02$. 
    Shown are four pairs of plots {\bf A}--{\bf D}, where {\bf A} and {\bf B} correspond to Hamiltonian $H_I$ (Eq.~\eqref{eq:MFIM}), and {\bf C} and {\bf D} correspond to Hamiltonian $H_H$ (Eq.~\eqref{eq:XXX}), while {\bf A} and {\bf C} correspond to single-site observable $O=Z_{j_0}$, and {\bf B} and {\bf D} correspond to two-site observable $O=Z_{j_0}Z_{j_0+1}$.
    In each pair, the left plot shows the difference $\abs{\mathcal{C}_A(t) - \mathcal{C}_W(t)}$ as a function of time $gt$, where $\mathcal{C}_W(t)$ is the autocorrelation function evaluated on the finite-window, with one-dimensional radial distance $R=6$, and $g$ is the Hamiltonian interaction strength. 
    It shows $\abs{\mathcal{C}_A(t) - \mathcal{C}_W(t)}$ for bond dimensions chosen such that the deviation lies well outside (blue), close to (orange), and within (green) the region bounded by $2\mathcal{B}_C(t)$.
    If the deviation lies within this bound, $\mathcal{C}_A(t)$ is at least as accurate as the approximation $\mathcal{C}_W(t)$.
    For {\bf A} and {\bf B}, $\chi=150$ lies entirely within this region, while for {\bf C} and {\bf D} $\chi=80$ and $\chi=70$ is sufficient, respectively.
    In each pair, the right plot shows $\mathcal{C}_A(t)$ as a function of time $gt$, with the bond dimension $\chi=150$.
    This is accompanied by a error bound $\pm \mathcal{B}_A(t)$ (Eq.~\eqref{eq:mpobound}), shown more clearly with the zoomed inset.
    }
    \label{fig:MPO_bound}
\end{figure*}

In the following results, we apply the framework to time-independent, non-integrable spin models, taking the finite-window Hamiltonian as a direct spatial restriction of the full system.
The autocorrelation function, {\it i.e.} with $A=B=O$ in Eq.~\eqref{eq:auto_generic} such that the evolved observable correlates with its initial value, is given by
\begin{equation}
    \mathcal{C}_j(t)= 2^{-n}\tr (O_j(t)O) \ .
    \label{eq:auto_function}
\end{equation}
With $m_A=m_B=m$, if the deviation $\Delta O(t)$ grows as $t^m$, then the autocorrelation deviation $\Delta \mathcal{C}(t) = \mathcal{C}_F(t) - \mathcal{C}_W(t)$ grows at least as $t^{2m}$. 
The autocorrelation function therefore provides a clear example in which the window approximation achieves a quadratic enhancement in accuracy relative to the underlying observable.
The following is thus based on the autocorrelation function.
Sec.~\ref{sec:1D} focuses on a one-dimensional spin chain, Sec.~\ref{sec:2D} focuses on the two-dimensional honeycomb lattice, while Sec.~\ref{sec:MPO} applies the framework to bounding other approximations, with MPO simulations as an illustrative example.

\subsection{One-dimensional system}

In one-dimension, we define the finite window as a chain of $2R+1$ spins embedded within the larger system, extending $R$ sites in either direction from a central site $j_0$.  
Since, in this case, operator growth from the initial observable site is restricted to two propagation directions, a window with a fixed number of spins is generally expected to yield the most accurate approximation in one-dimensional systems. 

In terms of Pauli $X$, $Y$, and $Z$ operators, two iconic non-integrable one-dimensional models are
\begin{align}
    H_I&= g\sum_{j=1}^{n-1} Z_jZ_{j+1} + h_x \sum_{j=1}^n X_j + h_z\sum_{j=1}^nZ_j 
    \label{eq:MFIM} \\
    H_H&= \frac{g}{3}\sum_{j=1}^{n-1}(X_jX_{j+1}+ Y_jY_{j+1} + Z_jZ_{j+1}) \ , \label{eq:XXX} 
\end{align}
where $H_I$ is the mixed-field Ising model with interaction strength $g$, and single-qubit field strengths $h_x$ and $h_z$~\cite{wurtz2020emergent,chiba2024proof,noh2021operator,craps2020lyapunov}.
$H_H$ is the isotropic Heisenberg model with interaction constant $g/3$~\cite{alba2015eigenstate,ljubotina2019kardar,denardis2020superdiffusion,dupont2021spatiotemporal}, such that the interaction strength of each Hamiltonian is normalized to the same local operator norm $g$.

To validate the strength of the bound $\mathcal{B}_C(t)$ (Eq.~\eqref{eq:correlationbound}), it is instructive to first consider the case of the Hamiltonian $H_I$ with $h_z=0$.
This corresponds to the transverse-field Ising model, where the exact autocorrelation function $\mathcal{C}_F(t)$ (Eq.~\eqref{eq:auto_function}) can be computed via the polynomially scaling Lie algebra generated by the terms $Z_jZ_{j+1}$ and $X_j$~\cite{orozco2024quantum,van2026fast,van2026quantum}, provided the observable $O$ is additionally an element of this Lie algebra.
In such cases, the error $\Delta \mathcal{C}(t)$ can be computed exactly.

This true deviation $\abs{\mathcal{C}_F(t) - \mathcal{C}_W(t)}$ is plotted in Fig.~\ref{fig:exact_auto}, as a function of time $gt$ in terms of the interaction strength $g$ of the Hamiltonian, for a chain of $n=50$ spins, approximated by a window of radius $R=6$, {\it i.e.} with 13 spins, for the single-site observable $O=X_{j_0}$ (panel ${\bf A}$) and two-site observable $O=Z_{j_0} Z_{j_0+1}$ (panel ${\bf B}$), with $j_0=25$.
Additionally plotted is the bound to the autocorrelation deviation in terms of the observable bound $\mathcal{B}_O(t)$ (Eq.~\eqref{eq:operatorbound}) in red, and the correlation bound $\mathcal{B}_C(t)$ (Eq.~\eqref{eq:correlationbound}) in blue.
The bound $\mathcal{B}_O(t)$ provides an extremely conservative bound to the true error, given that the autocorrelation deviation arises at an order at least twice as large as the underlying observable. 
The bound $\mathcal{B}_C(t)$, however, bounds the true error considerably more tightly, often with a value within an order of magnitude larger than the true error. 

With the confidence in the bound $\mathcal{B}_C(t)$, one can now consider the {\em non-integrable} cases, where the true deviation $\abs{\mathcal{C}_F(t) - \mathcal{C}_W(t)}$ is unobtainable. 
For the Hamiltonians $H_I$ (Eq.~\eqref{eq:MFIM}) and $H_H$ (Eq.~\eqref{eq:XXX}), the window autocorrelation function $\mathcal{C}_W(t)$ is considered for the single-site observable $O=Z_{j_0}$ and two-site observable $O=Z_{j_0}Z_{j_0+1}$.
In Fig.~\ref{fig:error_regions}, this is plotted as a function of time $gt$, for the window radii $R\in [1,6]$.
At short times, the results for all $R$ coincide, before the smaller windows begin to deviate from those with larger radii.
The shaded region, $\mathcal{C}_W(t)\pm \mathcal{B}_C(t)$, shown only while $\mathcal{B}_C(t)\le 0.3$, bounds the possible location of the full-system autocorrelation $\mathcal{C}_F(t)$.
As $R$ increases, the growth of this error region is delayed to later times, so that larger windows remain reliable for longer.
In particular, the largest window considered, $R=6$, provides an accurate approximation over a substantial time interval. 
This interval is slightly shorter for the two-site observable than for the single-site case, but in both cases it remains accurate considerably longer than what would be possible for the underlying observable.

\subsection{Two-dimensional system} 
\label{sec:2D}

The two-dimensional honeycomb spin system has appeared in contexts ranging from frustrated magnetism~\cite{gong2013phase,rehn2016classical} and quantum spin liquids~\cite{gong2013phase,rehn2016classical,liu2018dirac} to models of Kitaev materials~\cite{liu2018dirac,janssen2016honeycomb} and field-induced magnetic phases~\cite{liu2018dirac,janssen2016honeycomb,baek2017evidence}.
In the following, we consider the mixed-field Ising model within this geometry,
\begin{equation}
    H_I^{\text{2D}}= g\sum_{\left\langle i,j\right\rangle } Z_iZ_{j} + h_x \sum_{i} X_i + h_z\sum_{i}Z_i \ ,
    \label{eq:MFIM2D} 
\end{equation}
where $\left\langle i,j\right\rangle $ denotes nearest-neighbor pairs on the honeycomb lattice, and $g$ is the interaction strength.

Fig.~\ref{fig:honeycomb} shows three different finite-window configurations within the larger system, with the red spin indicating the central window site $j_0$.
Configuration {\bf A} extends one spin outward from the central site, while configuration {\bf B} extends two spins outward.
Three spins are included to configuration {\bf B} to form configuration {\bf C} comprising of three full hexagons, with the same number of spins as the $R=6$ radial distance window from the one-dimensional system.

Here, we consider the earliest times $gt$, in terms of the interaction strength of the Hamiltonian Eq.~\eqref{eq:MFIM2D}, at which the bounds $\mathcal{B}_O(t)$ (Eq.~\eqref{eq:operatorbound}) and $\mathcal{B}_C(t)$ (Eq.~\eqref{eq:correlationbound}) exceed $10^{-2}$.
For the observable $O=Z_{j_0}$, these times for $\mathcal{B}_O(t)$ are $gt=0.338$, $0.696$, and $0.742$, for configurations {\bf A}, {\bf B}, and {\bf C} respectively, while the corresponding times for $\mathcal{B}_C(t)$ are $gt=1.047$, $1.870$, and $2.011$.
As in the one-dimensional case, $\mathcal{B}_C(t)$ remains below the chosen threshold for substantially longer, demonstrating that it provides a considerably tighter bound than $\mathcal{B}_O(t)$.

Given that configurations {\bf A} and {\bf B} extend one and two spins outward to the window boundary from the central site respectively, they are therefore analogous to the one-dimensional windows with radial distances $R=1$ and $R=2$.
Consistent with this correspondence, these reported times are comparable to those obtained for the respective one-dimensional windows in Fig.~\ref{fig:error_regions} (top left panel), where the bound $\mathcal{B}_C(t)$ appears to exceed $10^{-2}$ at $gt\approx 1$ for $R=1$, and $gt\approx 2$ for $R=2$. 
Configuration {\bf C} does not uniformly extend the boundary to three spins from the observable, but increases the average radial distance to the window boundary relative to configuration {\bf B}. 
Accordingly, it yields a modest further improvement to the reported times.

\subsection{Bounding other approximations}
\label{sec:MPO}

While the full-system observable evolution and autocorrelation function can be approximated in various ways, many such methods do not generally provide a rigorous bound on their deviation from the exact dynamics. 
The bound (Eq.~\eqref{eq:correlationbound}) can, however, also be used to derive bounds for approximate expressions for autocorrelation functions based on approximations with no known bounds, as shown in the following.

Given the exact time-dependence $\mathcal{C}_F(t)$ of the autocorrelation function, the window approximation $\mathcal{C}_W(t)$ and another approximation $\mathcal{C}_A(t)$ of the autocorrelation function, the equality
\begin{equation} 
\mathcal{C}_F(t)-\mathcal{C}_A(t) = [\mathcal{C}_F(t)- \mathcal{C}_W(t)] + [\mathcal{C}_W(t)- \mathcal{C}_A(t)] \ ,
\end{equation}
yields the bound
\begin{equation} 
\abs{\mathcal{C}_F(t)-\mathcal{C}_A(t)} \le \mathcal{B}_C(t) + \abs{\mathcal{C}_W(t)- \mathcal{C}_A(t)}
=:
\mathcal{B}_A(t)\ ,
\label{eq:mpobound}\end{equation}
as direct consequence of the triangle inequality.

As an explicit example, the following is based on an MPO simulation $\mathcal{C}_A(t)$ with bond dimension $\chi$, to approximate the full-system autocorrelation function $\mathcal{C}_F(t)$. 
The deviation $\abs{\mathcal{C}_F(t)-\mathcal{C}_A(t)}$ is bounded by $\mathcal{B}_A(t)$. 
In particular, $\mathcal{C}_A(t)$ is evaluated for a chain of $n=51$ spins using TeNPy fourth order time-evolving block decimation (TEBD) with time-step $dt=0.02$, for the non-integrable one-dimensional Hamiltonians $H_I$ (Eq.~\eqref{eq:MFIM}) and $H_H$ (Eq.~\eqref{eq:XXX}).

As well as considering the error bound $\abs{\mathcal{C}_F(t)- \mathcal{C}_A(t)}\le \mathcal{B}_A(t)$ (Eq.~\eqref{eq:mpobound}), it is instructive to consider the bond dimension required to obtain an approximation of the full-system autocorrelation function that could be at least as accurate as $\mathcal{C}_W(t)$.
The exact autocorrelation is guaranteed to lie within a distance $\mathcal{B}_C(t)$ (Eq.~\eqref{eq:correlationbound}) of the finite-window result $\mathcal{C}_W(t)$.
The quantity $2\mathcal{B}_C(t)$ is the separation at which the midpoint between $\mathcal{C}_W(t)$ and $\mathcal{C}_A(t)$ just reaches this allowed region.
Thus if $\abs{\mathcal{C}_A(t) - \mathcal{C}_W(t)}\le 2\mathcal{B}_C(t)$ the MPO result could be at least as accurate as the finite-window result. 
If it lies outside this region, the MPO result is provably less accurate.

In Fig.~\ref{fig:MPO_bound}, this comparison is shown in the left-hand plot of panels {\bf A}--{\bf D}. 
For each case, the deviation $\abs{\mathcal{C}_A(t)-\mathcal{C}_W(t)}$ is taken relative to the $R=6$ window result $\mathcal{C}_W(t)$ and compared directly with the threshold $2\mathcal{B}_C(t)$.
Results are shown only up to the first time $gt$ at which $\mathcal{B}_C(t)$ exceeds $10^{-2}$, restricting the comparison to the regime in which the window result remains tightly bounded. 
Panels {\bf A} and {\bf B} correspond to the Hamiltonian $H_I$ (Eq.~\eqref{eq:MFIM}), while panels {\bf C} and {\bf D} correspond to $H_H$ (Eq.~\eqref{eq:XXX}). 
Panels {\bf A} and {\bf C} consider the single-site observable $O=Z_{j_0}$, whereas panels {\bf B} and {\bf D} consider the two-site observable $O=Z_{j_0}Z_{j_0+1}$, with $j_0$ denoting the central site of the spin chain. 
For each panel, three bond dimensions are chosen to show the increasing accuracy of the MPO approximation; one for which $\abs{\mathcal{C}_A(t)-\mathcal{C}_W(t)}$ remains outside the region defined by $2\mathcal{B}_C(t)$ (blue), one for which it lies close to the boundary (orange), and one for which it lies within this region (green).
In the latter case, the MPO approximation could be at least as accurate as the finite-window result.
By this criteria, bond dimension $\chi=150$ is sufficient for panels {\bf A} and {\bf B}, while $\chi=80$ and $\chi=70$ are sufficient for panels {\bf C} and {\bf D}, respectively.

The right plot of these panels shows $\mathcal{C}_A(t)$ as a function of time $gt$ for bond dimension $\chi=150$, again restricted to times before $\mathcal{B}_C(t)$ exceeds $10^{-2}$.
The corresponding error region $\mathcal{C}_A(t)\pm \mathcal{B}_A(t)$ (Eq.~\eqref{eq:mpobound}) is additionally shown, and highlighted with a zoomed inset.
This is a tight error region, reflecting the similarly tight bound $\mathcal{B}_C(t)$ provided by the finite-window approximation.

\section{Discussion}

The framework and results outlined in this work extend our ability to provide classically simulated approximations of a quantum system's dynamics, with guaranteed accuracy, beyond integrable or small systems.
While quantum technology develops with the goal to eventually replace the classical simulation of quantum dynamics with the actual quantum simulation, the classically simulated result can be essential for benchmarking the quantum simulation.
Useful and computable bounds to the approximation error, like those laid out in this work, are thus central to the practicability of classical simulation.

The finite-window approximation provides controlled access to dynamical observables in the full many-body system, including geometries beyond one dimension, and enables an accurate approximation to the infinite-temperature correlation function with an accompanying tight error bound.
Thus any quantity that is evaluated in terms of the approximated correlation function can be additionally bounded.
While we apply this framework to correlation functions, there are several other important quantities that are evaluated in terms of dynamical observables, such as out-of-time-ordered commutators (OTOCs)~\cite{seki2025simulating,tanner2025learning,jonay2025two,ozaki2025disorder} and full counting statistics (FCS)~\cite{valli2025efficient,joshi2025measuring,horvath2026full,landi2024current}, that would likely also benefit from the methodology outlined in this work.

The application of the framework to other important approximations, such as MPO simulations, establishes error bounds for methods that typically lack them.
This suggests a powerful and general route for combining local exact calculations with scalable approximate methods.

Broadly, these results establish finite-window dynamics as a practical route to provably accurate classical simulation of many-body quantum dynamics, linking exact local calculations and scalable approximations in regimes where the full-system dynamics are out of reach.

\section*{Acknowledgements}
This work was supported by the U.K. Engineering and Physical Sciences Research Council via the studentship (EPSRC DTP - EP/W524323/1) and
via the EPSRC Hub in
Quantum Computing and Simulation (EP/W524311/1).
Numerical simulation routines were performed on the Imperial HPC cluster.

\section*{Data availability}
The code used to generate the data used in this paper is available without restriction~\cite{van_lomwel_code}.

\bibliography{auto_library.bib}

\end{document}